\documentclass{ws-procs9x6-cpt16}
\begin{document}

\newcommand{\refeq}[1]{(\ref{#1})}
\def\etal {{\it et al.}}

\title{Preliminary Design of a Pendulum Experiment for\\
Searching for a Lorentz-Violation Signal}

\author{Cheng-Gang Shao, Ya-Fen Chen, and Yu-Jie Tan}

\address{MOE Key Laboratory of Fundamental Physical Quantities Measurement\\
School of Physics, Huazhong University of Science and Technology\\
1037 Luo Yu Road, Wuhan 430074, People's Republic of China}

\begin{abstract}
This work mainly presents a preliminary design for a pendulum experiment 
with both the source mass and the test mass in a striped pattern
to amplify the Lorentz-violation signal, 
since the signal is sensitive to edge effects.
\end{abstract}

\bodymatter

\section{Introduction}
Studying violations of local Lorentz invariance in the spacetime theory of gravity is a way to probe General Relativity.
To search for Lorentz violation in pure gravity, some short-range experiments\cite{1,2,3}
such as testing the gravitational inverse-square law using a torsion pendulum
were performed and the data analyzed for a Lorentz-violation signal.
In this work, we mainly present a special design
for a pendulum experiment to enhance the sensitivity to the violation signal,
which has not yet been detected.

\section{Starting point of the new design}
Lorentz-invariance violation in the pure-gravity sector with quadratic couplings of the Riemann curvature has been described quantitatively
by the Standard-Model Extension (SME).\cite{1,4} Using the effective field technique, the coupling introduces a correction for the Newton gravity,
which depends on time and orientation. Corresponding to the gravitational potential, the correction can be written as:
\begin{equation}\label{Eq1}
{V_{LV}}(\vec{x})
= - G \frac{{m_1}{m_2}}{|\vec{x}|^{3}}{\bar k}(\hat{x},T),
\end{equation}
with
\begin{equation}\label{Eq2}
{\bar k}(\hat{x},T)\equiv \!\!\tfrac{3}{2}{({{\bar k}_{\rm{eff}}})_{jkjk}}-9{({{\bar k}_{\rm{eff}}})_{jkll}}\hat{x}^{j}\hat{x}^{k}
+\tfrac{15}{2}{({{\bar k}_{\rm{eff}}})_{jklm}}
\hat{x}^{j}\hat{x}^{k}\hat{x}^{l}\hat{x}^{m}.
\end{equation}
Here, $m_{1}$ and $m_{2}$ represent two point masses, and $\vec{x}$ is the separation between them. 
The quantity $\hat{x}^{j}$
is the projection of the unit vector $\vec{x}$ along the $j$th direction. 
The coefficient ${({{\bar k}_{\rm{eff}}})_{jklm}}$
for Lorentz violation 
has 15 independent components, 
as it is totally symmetric with indices $j$, $k$, $l$, $m$
ranging over the three spatial directions.

As discussed in Ref.\ \refcite{5}, shape and edge effects play an important role in determining the sensitivity of the
experiment to the coefficients for Lorentz-invariance violation:
\begin{equation}\label{Eq3}
\tau_{LV} \sim\varepsilon \Delta C{({\bar k_{\rm{eff}}})_{jklm}},
\end{equation}
where $\varepsilon$ represents the edge effect, which is related to the geometrical parameters of the test and source masses. To
derive an amplified experimental sensitivity to the Lorentz-violation signal, $\varepsilon$ should be designed as large as possible.
However, for our pendulum experiment, we have to work within the maximum capability of the fiber.

\section{Theoretical analysis for the new design}
According to Eq.\ (\ref{Eq2}), the violation coefficients ${({{\bar k}_{\rm{eff}}})_{jklm}}$ are different in different frames. Thus,
the Sun-centered frame is usually adopted as the canonical frame to report the results from experiments searching for a Lorentz-violation signal, 
since the coefficients can be regarded as constant 
on the scale of the solar system. The violation coefficients
${({{\bar k}_{\rm{eff}}})_{JKLM}}$ in the Sun-centered frame can be connected with the coefficients
${({{\bar k}_{\rm{eff}}})_{jklm}}$ in the laboratory frame by the rotation matrix $R^{jJ}$:
\begin{equation}\label{Eq4}
({{\bar k}_{\rm{eff}}})_{jklm}=R^{jJ}R^{kK}R^{lL}R^{mM}({{\bar k}_{\rm{eff}}})_{JKLM},
\end{equation}
where $R^{jJ}$ involves $\omega_{\oplus}\!\!\simeq \!\!2\pi/(\rm{23~h~56~min})$. Thus, Eq.\ (\ref{Eq2}) can be expressed as a Fourier series
in the sidereal time $T$ as:
\begin{equation}\label{Eq5}
{\bar k}(\hat{x},T) =\! {c_0} + \sum\limits_{m = 1}^4 \left[ {c_m}\cos (m{\omega _ \oplus }T) + {s_m}\sin (m{\omega _ \oplus }T) \right]
\end{equation}
through Eqs.\ (\ref{Eq2}) and (\ref{Eq4}). The nine Fourier amplitudes ($c_0$, $c_m$, $s_m$) are functions of $({{\bar k}_{\rm{eff}}})_{JKLM}$.
  \begin{figure}[!t]
\includegraphics[width=3.4in]{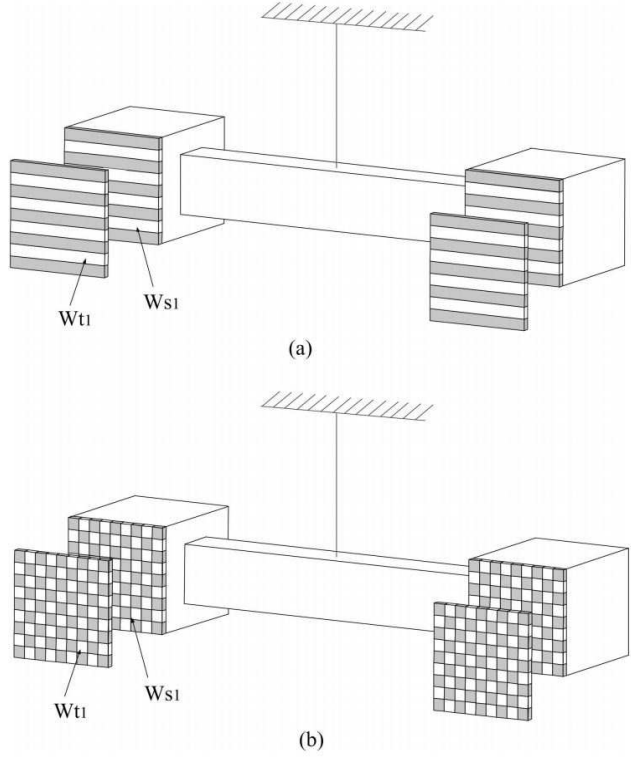}
\caption{The I-shape torsion pendulum. $\rm{W}_{t1}$ and $\rm{W}_{s1}$ represent the test mass and source mass, respectively.
 The masses are in the striped geometry (a) and in the checkered geometry (b).
}
\label{aba:fig1}
\end{figure}
We reassemble the $({{\bar k}_{\rm{eff}}})_{JKLM}$ (15 dimensions), and redefine the violation coefficients as ${\bar k}_{j}$ (15 dimensions).
By introducing ${\bar k}_{j}$, the space $({{\bar k}_{\rm{eff}}})_{JKLM}$ is decomposed into 5 subspaces, in which different harmonics of
$\omega_{\oplus}$ for the Lorentz-violation signal are linked to different subspaces, such as:
\begin{equation}\label{Eq6}
{c_0} = {\alpha _0}{\bar k_0} + {\alpha _1}{\bar k_1} + {\alpha _2}{\bar k_2}.
\end{equation}
We find ${c_0}$ is linked to $k_0$, $k_1$, $k_2$, and the other four harmonics of $\omega_{\oplus}$ correspond to the other four subspaces, respectively. Thus, for the
Lorentz-violation torque,
\begin{eqnarray}\label{Eq7}
  {\tau _{LV}} \!\!=&& \!\!\!G{\rho _1}{\rho _2}\!\!\iint {d{V_1}d{V_2}}\frac{\partial }{{\partial \theta }}\frac{{\bar k(\hat r,T)}}{{{r^3}}} \hfill  \nonumber\\
  {\text{     }} =&&\!\!\! {\Lambda _j} {\bar k}_{j}.
\end{eqnarray}
 Here, the transfer coefficient ${\Lambda _j}$ is related to ${\alpha _j}$ and the geometrical parameters of the test mass and source mass, and it
includes edge effects as described by $\varepsilon$ in Eq.\ (\ref{Eq3}). According to Eqs.\ (\ref{Eq5})-(\ref{Eq7}), one can seek a special design for the experiment
 to satisfy the particular research requirements. For example, to probe the Lorentz-violation coefficients ${\bar k}_{0}$, ${\bar k}_{1}$ and ${\bar k}_{3}$ better,
  one can design the test mass and source mass to make ${\Lambda _0}$, ${\Lambda _1}$ and ${\Lambda _2}$ larger.
\section{Experimental design}
The experimental schematic is similar to that in testing the inverse-square law for HUST-2011 (see Fig.\ 2 in Ref.\ \refcite{5}), but
the geometrical parameters here are designed differently to amplify the violation signal. We analyzed two upgraded designs (see Fig.\ 1) for the torsion pendulum.
In one, the shape of the masses involves a striped geometry (see Fig.\ 1(a)), and in the other it involves a checkered geometry (see Fig.\ 1(b)).
For the test mass $\rm{W}_{t1}$ and the source mass $\rm{W}_{s1}$ in the two designs, the gray part and white part represent tungsten and 
glass, respectively. Comparing the results of numerical simulation, we find the first design (striped geometry) is the better option to increase 
the Lorentz-violation signal. 

\section{Summary}
Theoretically, we decomposed the 15 Lorentz-violation coefficients into five parts,
with different harmonics of the violation signal corresponding to different parts, which
helps to perform the special experimental design required to study a certain violation coefficient. In addition, we proposed a design to search for Lorentz violation at higher sensitivity,
in which the masses are in a striped pattern.

\section*{Acknowledgments}
This work was supported by the National Natural Science Foundation of China (11275075).

\end{document}